\documentclass[aps,prl,reprint,showpacs,superscript]{revtex4-1}
\usepackage{graphicx}
\begin{document}
\newcommand{\im}[0]
{{\rm i}}
\newcommand{\ex}[0]
{{\rm e}}
\newcommand{\ket}[1]
{|#1 \rangle}
\newcommand{\bra}[1]
{\langle #1 |}
\newcommand{\vect}[1]
{\mbox{{\boldmath $#1$}}}
\newcommand{\svect}[1]
{{\mbox{{\scriptsize \boldmath $#1$}}}}
\newcommand{\singlet}[0]
{{}^{1}S_{0}}
\newcommand{\triplet}[0]
{{}^{3}P_{2}}
\newcommand{\vlong}[0]
{V_{\rm long}}
\newcommand{\vshort}[0]
{V_{\rm short}}
\newcommand{\vdiag}[0]
{V_{\rm diag}}
\newcommand{\slong}[0]
{s_{\rm long}}
\newcommand{\sshort}[0]
{s_{\rm short}}
\newcommand{\sdiag}[0]
{s_{\rm diag}}
\newcommand{\er}[0]
{E_{\rm R}^{(1064)}}
\newcommand{\ers}[0]
{E_{\rm R}^{(532)}}
\newcommand{\kl}[0]
{k_{\rm L}}

\title{Coherent driving and freezing of bosonic matter wave in an optical Lieb lattice}
\author{Shintaro Taie}
\altaffiliation{Electronic address: taie@scphys.kyoto-u.ac.jp}
\affiliation{Department of Physics, Graduate School of Science, Kyoto University, Japan 606-8502}
\author{Hideki Ozawa}
\affiliation{Department of Physics, Graduate School of Science, Kyoto University, Japan 606-8502}
\author{Tomohiro Ichinose}
\affiliation{Department of Physics, Graduate School of Science, Kyoto University, Japan 606-8502}
\author{Takuei Nishio}
\affiliation{Department of Physics, Graduate School of Science, Kyoto University, Japan 606-8502}
\author{Shuta Nakajima}
\affiliation{Department of Physics, Graduate School of Science, Kyoto University, Japan 606-8502}
\author{Yoshiro Takahashi}
\affiliation{Department of Physics, Graduate School of Science, Kyoto University, Japan 606-8502}
\date{\today}
\begin{abstract}
		While kinetic energy of a massive particle generally has quadratic dependence on its momentum,
		a flat, dispersionless energy band is realized in crystals with specific lattice structures.
		Such macroscopic degeneracy causes the emergence of localized eigenstates and has been a key concept
		in the context of itinerant ferromagnetism.
		Here we report the realization of a ``Lieb lattice'' configuration with an optical lattice, which has a flat energy band
		as the first excited state. Our optical lattice potential possesses various degrees of freedom about its manipulation,
		which enables coherent transfer of a Bose-Einstein condensate into the flat band.
		In addition to measuring lifetime of the flat band population for different tight-binding parameters,
		we investigate the inter-sublattice dynamics of the system by projecting the sublattice population onto
		the band population. This measurement clearly shows the formation of the localized state with the specific
		sublattice decoupled in the flat band, and even detects the presence of flat-band breaking perturbations, resulting
		in the delocalization.
		Our results will open up the possibilities of exploring physics of flat band with a highly controllable quantum system.
\end{abstract}
\maketitle
Many-body properties of a quantum system show drastic changes according to the geometry of an underlying
lattice structure. One of the textbook examples is an antiferromagnet on a frustrated lattice geometry
\cite{Balents2010}, where the geometric frustration prevents spins from N\`{e}el ordering and the system
exhibits more nontrivial, correlated ground state. Specific lattice geometry
can also induce frustration of kinetic energy. In this case a macroscopic number of momentum eigenstates are
energetically degenerate, forming a dispersionless flat band. Flat bands have been playing an important role
in theoretical study of itinerant ferromagnetism, as the presence of interaction lifts the bulk degeneracy
and chooses the ferromagnetic ground state \cite{Lieb1989,Mielke1991,Tasaki1992}.

\begin{figure}[bt]
	\includegraphics[width=85mm]{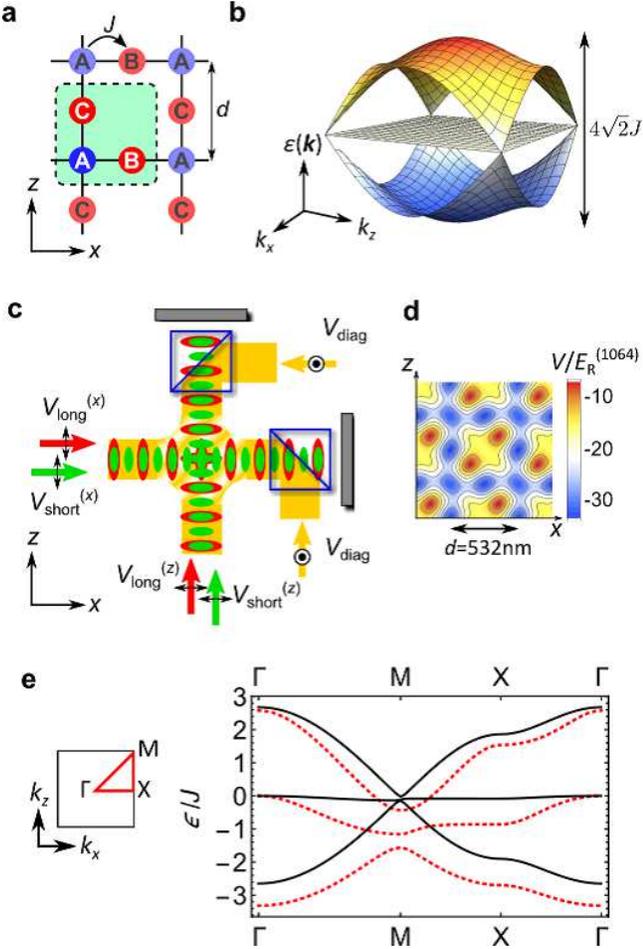}
	\caption{
		\textbf{Optical Lieb lattice.}
		\textbf{a}, Lieb lattice. A unit cell is indicated by the green square.
		\textbf{b}, Tight-binding energy band structure of the Lieb lattice.
		\textbf{c}, Experimental realization of the Lieb lattice. Black arrows indicate polarizations of the lattice beams.
		\textbf{d}, Lattice potential for $(\slong, \sshort, \sdiag) = (8, 8, 9.5)$ at $\phi_x = \phi_z =0$ and $\psi=\pi/2$.
		\textbf{e}, Band structures of the optical Lieb lattice at: $(\slong, \sshort, \sdiag) = (8, 8, 9.5)$ (red dashed),
		$(34, 34, 37.4)$ (black solid).
	}
	\label{fig_setup}
\end{figure}

Ultracold atomic gases in a periodic potential generated by standing waves of laser light (optical lattices) have made
a great success in realizing controllable many-body systems described by well-defined theoretical models of interest,
such as the Hubbard model \cite{Bloch2008,Esslinger2010}.
Recently, increasing experimental efforts have been made to create and investigate non-standard
(other than simple cubic) optical lattices which have unique geometric features
\cite{Becker2010,Olschlager2011,Wirth2011,SoltanPanahi2011,Struck2011,SoltanPanahi2012,Tarruell2012,Jo2012,Windpassinger2013}.
Among them, a kagome lattice \cite{Jo2012} is a promising candidate for exploring flat band physics.
However, the negative hopping amplitude, which is natural for ultracold atoms, forces the flat band
to be energetically highest among the three $s$-orbitals, and so far phenomena characteristic to the flat band
have not been reported.
A different type of lattice structure known as a Lieb lattice, also referred as a decorated square lattice,
or a $d$-$p$ model in the analogy to CuO${}_2$ plane of a High-$T_{\rm c}$ superconductor \cite{Iglovikov2014},
similarly has a flat band as the second (first excited) band.
It consists of two sublattices:
one of them forms a standard square lattice (the $A$-sublattice in Fig.~\ref{fig_setup}a) and the other lies on
every side of the square. For convenience, we further divide the latter into the $B$- and $C$-
sublattices. The single particle energy spectrum in the tight binding limit (Fig.~\ref{fig_setup}b)
has the characteristic flat band and the Dirac cone on the corner of the Brillouin zone. 
This lattice satisfies the criteria for the occurrence of Lieb's ferrimagnetism, which states that the half-filled
spin-$1/2$ fermions exhibits nonzero magnetization for a positive on-site interaction \cite{Lieb1989}.
Also for bosonic systems, a flat band gives a fascinating question of whether condensation is possible
in the presence of kinetic energy frustration. Theoretical investigation predicts supersolid order for
a flat band \cite{Huber2010}.
The Lieb lattice was recently realized in a photonic lattice \cite{Guzman2014} where the electric field
passing through the lattice emulates time evolution of single-particle wave function.
Also, Bose condensation of polaritons in one-dimensional analogue of the Lieb lattice (sawtooth lattice)
has also been reported \cite{Baboux2015}.
Optical-lattice realization of the Lieb lattice provides powerful quantum simulation of a closed, interacting
many-body system.

 In this paper, we present manipulation and detection of bosonic matter-wave
in an optical Lieb lattice. An atomic condensate is coherently loaded into the flat band and its time evolution is
measured in two ways: quasimonentum- and sublattice-resolved detection, which reveal the localized
character of the flat band wave function.
Relatively short lifetime of atoms in the 2nd band was observed, while it can be made longer by
increasing the band gap to the lowest band.
This work paves the way to experimental study of flat band physics with cold atoms.
Using Fermi gases with the Fermi energy lying at the flat band can avoid the lifetime problem and
will provide ideal playground for investigating flat band ferromagnetism \cite{Noda2009,Noda2014,Chen2014b}
and topological phases with artificial gauge fields \cite{Goldman2011}.

\section{Formation of an optical Lieb lattice}
We construct the optical Lieb lattice by superimposing three types of optical lattices
(see Fig.~\ref{fig_setup}c and \ref{fig_setup}d), leading to the potential
\begin{eqnarray}
	V(x, z) = &-& \vlong^{(x)}  \cos^2 (\kl x) -\vlong^{(z)} \cos^2 (\kl z) \nonumber \\
					&-& \vshort^{(x)}  \cos^2 (2\kl x+\phi_x) -\vshort^{(z)} \cos^2 (2\kl z+\phi_z) \nonumber \\
					&-& \vdiag \cos^2 \left[ \kl \left( x-z \right) + \psi \right], \label{eq_potential}
\end{eqnarray}
where $z$ indicates the direction of gravity.
Here, $\kl = 2\pi/\lambda$ is a wavenumber of a long lattice (with a depth $\vlong$), for which we choose
$\lambda = 1064$~nm. A short lattice ($\vshort$) is formed by laser beams at $532$~nm.
A diagonal lattice ($\vdiag$) with the wave number $\sqrt{2}\kl$ is realized by interference of the
mutually orthogonal laser beams at $532$~nm along the $x$- and $z$-directions.
Compared to the proposals \cite{Shen2010,Apaja2010}, equation~(\ref{eq_potential}) lacks a diagonal lattice
with $x+z$ spatial dependence. While this causes slightly larger discrepancy from the ideal tight-binding energy bands,
the sufficiently deep lattices can reproduce the desired energy spectrum including the flat band, as seen in Fig.~\ref{fig_setup}e.
In the following, we specify each lattice depth in unit of long-lattice recoil energy as
$(\slong, \sshort, \sdiag) = (\vlong, \vshort, \vdiag)/\er$, where $\er=\hbar^2 \kl^2/(2m)$ and
$m$ is the atomic mass of $^{174}$Yb.
The motion of free bosons in the $x$-$z$ plane is governed by tight-binding Hamiltonian
\begin{eqnarray}
	\hat{H} = -J\sum_{\langle i,j \rangle} \left( \hat{a}^{\dagger}_i \hat{a}_j + \text{H.c.} \right)
	+ \sum_{S=A,B,C} E_S \sum_{i \in S}\hat{n}_i,
	\label{eq_Hamiltonian}
\end{eqnarray}
where $\hat{a}_i$ is the annihilation operator on site $i$ and $\hat{n}_i = \hat{a}^\dagger_i \hat{a}_i$.
The nearest-neighbor hopping amplitude $J$ is mainly determined by $\vshort$, whereas the other lattice depths
set an energy offset $E_S$ of each sublattice. Excepting the contribution from the zero-point energies of
each potential well, they are approximately given by
$E_A \sim -\vlong^{(x)}-\vlong^{(z)}$, $E_B \sim -\vlong^{(z)}-\vdiag$, and $E_C \sim -\vlong^{(x)}-\vdiag$,
which are able to be independently controlled by tuning the lattice depths.
In the $y$-direction which is perpendicular to the Lieb lattice plane, the atoms are weakly confined
in a harmonic trap (one-dimensional tube configuration), unless otherwise specified.
In addition to each lattice depth, three phase parameters should be set to $\phi_x = \phi_z =0$ and $\psi=\pi/2$
in order to realize the characteristic three-sublattice structure of the Lieb lattice depicted in Fig.~\ref{fig_setup}d (Methods).

\section{Loading BEC into a flat band by phase imprinting}
\begin{figure}[bt]
	\includegraphics[width=85mm]{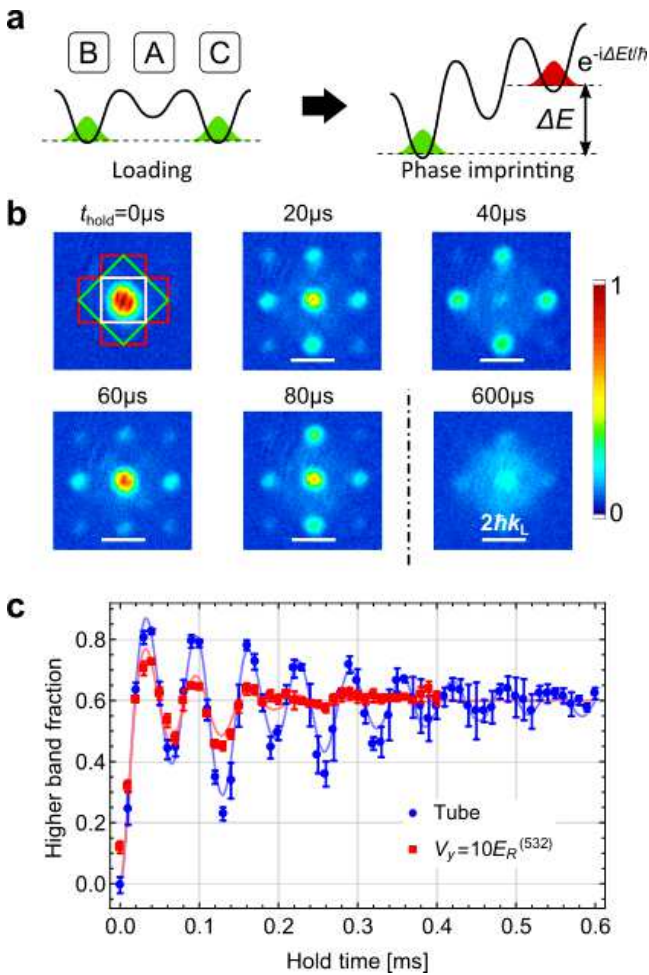}
	\caption{
		\textbf{Coherent band transfer.}
		\textbf{a}, Principle of the transferring method.
		\textbf{b}, Absorption images reveal the coherent oscillations between the $\ket{B}+\ket{C}$ and
		$\ket{B}-\ket{C}$ states.
		In the upper left image, the first three Brollouin zones are displayed by white, green, and red
		lines, respectively.
		\textbf{c}, Oscillating behavior of the band population during phase imprinting in absence of
		lattice confinement along the $y$-direction  (blue circles), and with lattice confinement $-V_y \cos^2 (2\kl y)$
		(red squares).
		Solid lines are the fit results using the single particle solution of the Schr\"{o}dinger equations (see main text).
		Error bars denote standard deviation of three independent measurements.
	}
	\label{fig_bandtransfer}
\end{figure}

One of the fundamental properties of flat bands is the localization of the wave function as a consequence of
quantum-mechanical interference of traveling matter waves. The localization is due to a purely
geometric effect, as we briefly explain below.
The Hilbert space for the Lieb lattice in the tight-binding regime is spanned by the quasimomentum eigenstates
of each sublattice $\ket{\vect{k},S}$ $(S=A, B, C)$.
Nearest neighbor tunneling induces the momentum dependent coupling $\mathcal{J}_{AB} = J \cos{k_x d/2}$
between the $A$- and $B$-sublattices, and similarly $\mathcal{J}_{AC}=J \cos{k_z d/2}$ between
the $A$- and $C$-sublattices, where $d=532$~nm is the lattice periodicity.
The flat band states are the zero-energy eigenstates $\cos \theta \ket{\vect{k},B} - \sin \theta \ket{\vect{k},C}$ with
$\tan \theta = \mathcal{J}_{AB}/\mathcal{J}_{AC}$, which have no amplitude on the $A$-sublattice.
Consequently, a wave packet composed of the flat band states remains localized, as the tunneling from
a $B$-site and a $C$-site destructively interfere on the adjacent $A$-site.
We explore this nature in the following experiments.
Here we note that the flat band in the Lieb lattice is mathematically equivalent to ``dark states'' of
laser-coupled $\Lambda$-type three level systems in atomic physics. Here, three sublattices correspond to the basis
of three levels, tunneling amplitudes serve as laser-induced coupling, and energy difference of each sublattice plays a role of
the detuning of laser.

In a Lieb lattice, a flat band is realized as the first excited band, hence a BEC loaded adiabatically into
an optical Lieb lattice is not populated in the flat band. However, tunability of our optical Lieb lattice
enables to coherently transfer the population in the lowest band into the flat band by phase imprinting
(see Fig.~\ref{fig_bandtransfer}a).
The scheme is easily understood by considering tight-binding wave functions in each band.
At zero quasimomentum and in the equal-offset condition $E_A = E_B = E_C$, a simple calculation gives
$\ket{\text{1st}} = \ket{A}+(\ket{B}+\ket{C})/\sqrt{2}$, $\ket{\text{2nd}} = \ket{B}-\ket{C}$,
and $\ket{\text{3rd}} = \ket{A}-(\ket{B}+\ket{C})/\sqrt{2}$ from the 1st to the 3rd band,
where we omit the momentum indices from the sublattice eigenstates (see also Supplementary Information).
Taking advantage of rich controllability in our lattice potential, we can smoothly modify these eigenstates.
With sufficiently large $\vdiag$ (equivalently with large $E_A-E_{B,C}$), the lowest Bloch state has
essentially no amplitude in the $A$-sublattice, allowing to realize $\ket{B}+\ket{C}$ state.
Next, we apply sudden change in one of the long lattice, $\vlong^{(z)}$. This creates the energy difference
between the $B$- and $C$-sublattices and the relative phase of the condensate wave function starts to evolve
with a period $2\pi \hbar/(E_C-E_B)$.
In the basis of the initial band structure, this time evolution is a coherent oscillation between
$\ket{\text{1st}}$ and $\ket{\text{2nd}}$.

The explicit procedure of the loading and detecting a condensate in the flat band is as follows.
We adiabatically load a BEC of $2 \times 10^4$ ${}^{174}$Yb atoms into the Lieb lattice with
$(\slong, \sshort, \sdiag) = (8, 8, 20)$ and apply sudden increase of $\slong^{(z)}$ to $26.4$ for variable duration.
At the same time, we ramp $\sshort$ up to $20$ to prevent tunneling during the band transfer.
After this sequence, we return the lattice depths to the initial values and perform adiabatic turning off of the lattice potential
in order to map quasimomentum to free-particle momentum (band mapping) \cite{Greiner2001,Koehl2005}.
Figure~\ref{fig_bandtransfer}b shows the absorption images taken after $14$~ms of the ballistic expansion,
which reveal the interband dynamics of a condensate. At zero quasimomentum, atoms in the 2nd and 3rd band
are mapped to the same point of the Brillouin zone. In addition, the finite spread of the condensate makes
it difficult to precisely distinguish the population in the 2nd Brillouin zone from other neighboring zones.
Therefore, Instead of plotting the population in the 2nd Brillouin zone, here we count atoms in the 1st Brillouin zone,
and show the fraction of atoms in the other higher bands in Fig.~\ref{fig_bandtransfer}c. For momentum space analysis,
see Section III of Supplementary Information.
We fit the data with the function in the form $a \exp (-t/\tau) F_{\rm th}(t) + b(1-\exp(-t/\tau))$, where $F_{\rm th}(t)$
is the numerical solution of the single particle Schr\"{o}dinger equation. In fitting the data, we adopt
$\slong^{(z)}$ during the band transfer as a free parameter, and obtain the best fit with $\slong^{(z)}=25.5$,
close to the expected value of $26.4$. Although the oscillations involve non-negligible contributions from the higher bands,
at the half period of the first cycle we expect that $> 75$\% of atoms are transferred to the 2nd band. 
This transfer method is also applicable in the presence of the lattice confinement along the $y$-axis,
though the decay time of the oscillation $\tau =86(7)$~$\mu$s is much shorter than the case of a weak harmonic
confinement (1D-tube), $\tau =260(10)$~$\mu$s.

\section{Relaxation dynamics of a flat band}
\begin{figure*}[hbt]
	\includegraphics[width=160mm]{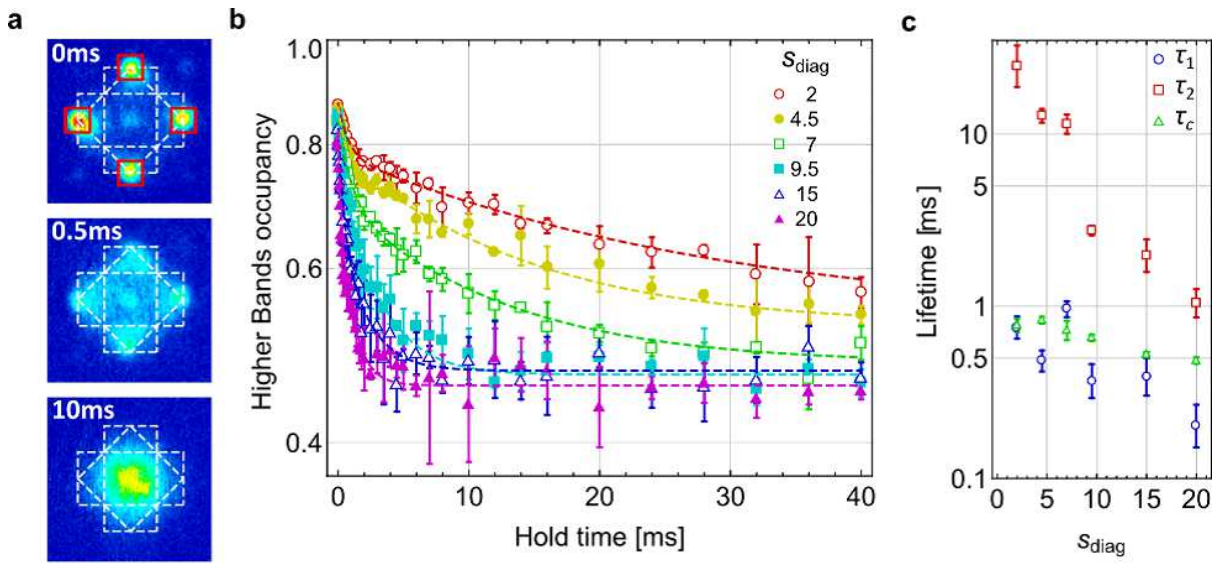}
	\caption{
		\textbf{Lifetime of atoms in the flat band.}
		\textbf{a}, Absorption images for the lifetime measurement of the 2nd band with three different hold times,
		taken after $14$~ms time-of-flight. The diagonal lattice depth is $\sdiag=9.5$.
		The first three Brillouin zones are indicated by the white dashed lines.
		In the top image, the areas used to evaluate the lifetime of a condensate ($\tau_{\rm c}$) are also displayed
		with the red squares.
		\textbf{b}, Decay of the flat band at $(\slong, \sshort)= (8, 8)$ and variable $\sdiag$. Solid lines are
		the fit results with double exponential curves. Error bars denote the standard deviation of three independent
		measurements.
		\textbf{c}, Lifetime of the flat band. $\tau_{1,2}$ are the fast and slow decay time obtained from the data
		shown in \textbf{b}, respectively. $\tau_{\rm c}$ is the ${\rm e}^{-1}$ lifetime of condensates. 
		Error bars represent fitting error.
	}
	\label{fig_lifetime}
\end{figure*}

We measure the lifetime of atoms in the 2nd band of the optical Lieb lattice.
After transferring to the 2nd band, we change the depth $\sdiag$ of the diagonal lattice to control the energy gap
between the 1st and 2nd band. As well as the band gap \cite{Mueller2007}, lifetime of a quantum gas
in the excited band is strongly affected by the density overlap with the states in the lower bands \cite{Olschlager2012}.
As we increase $\sdiag$, the average gap between the 1st and 2nd band becomes smaller and at the same time
their density profiles become similar to each other. In the opposite limit of shallow $\sdiag$, the band gap increases and two bands
have no density overlap, as the lowest band mostly consists of the $A$-sublattice.
We take a variable hold time in the lattice, followed by band mapping to count the atom number in the excited bands.
Typical absorption images are shown in Fig.~\ref{fig_lifetime}a.
The decay curves displayed in Fig.~\ref{fig_lifetime}b show expected behavior of increasing lifetime with decreasing $\sdiag$.
In addition, increasing the gap makes the dynamics more clearly separate into two process: decay of the condensate
within the 2nd band (middle image of Fig.~\ref{fig_lifetime}a) and decay of atoms into the lowest band (bottom image).
We find that the curve is well fitted by double exponential with the form $a_1 \exp(-t/\tau_1)+a_2 \exp(-t/\tau_2)+b$.
The fast component $\tau_1$ shows only weak dependence on $\sdiag$, whereas the slow component $\tau_2$ shows
over $20$-fold changes from the smallest to largest $\sdiag$. We also extract the lifetime of the condensate
in the 2nd band, $\tau_{\rm c}$, by counting atoms on the corner of the 2nd Brillouin zone (Figure~\ref{fig_lifetime}a),
and find similar behavior with $\tau_1$. This implies that the intial fast decay is related to the decay of the condensate,
which involves the decay to the lower band with faster time constant compared with the non-condensed atoms.

\section{Localization of a wave function in a flat band}
\begin{figure*}[bt]
	\includegraphics[width=160mm]{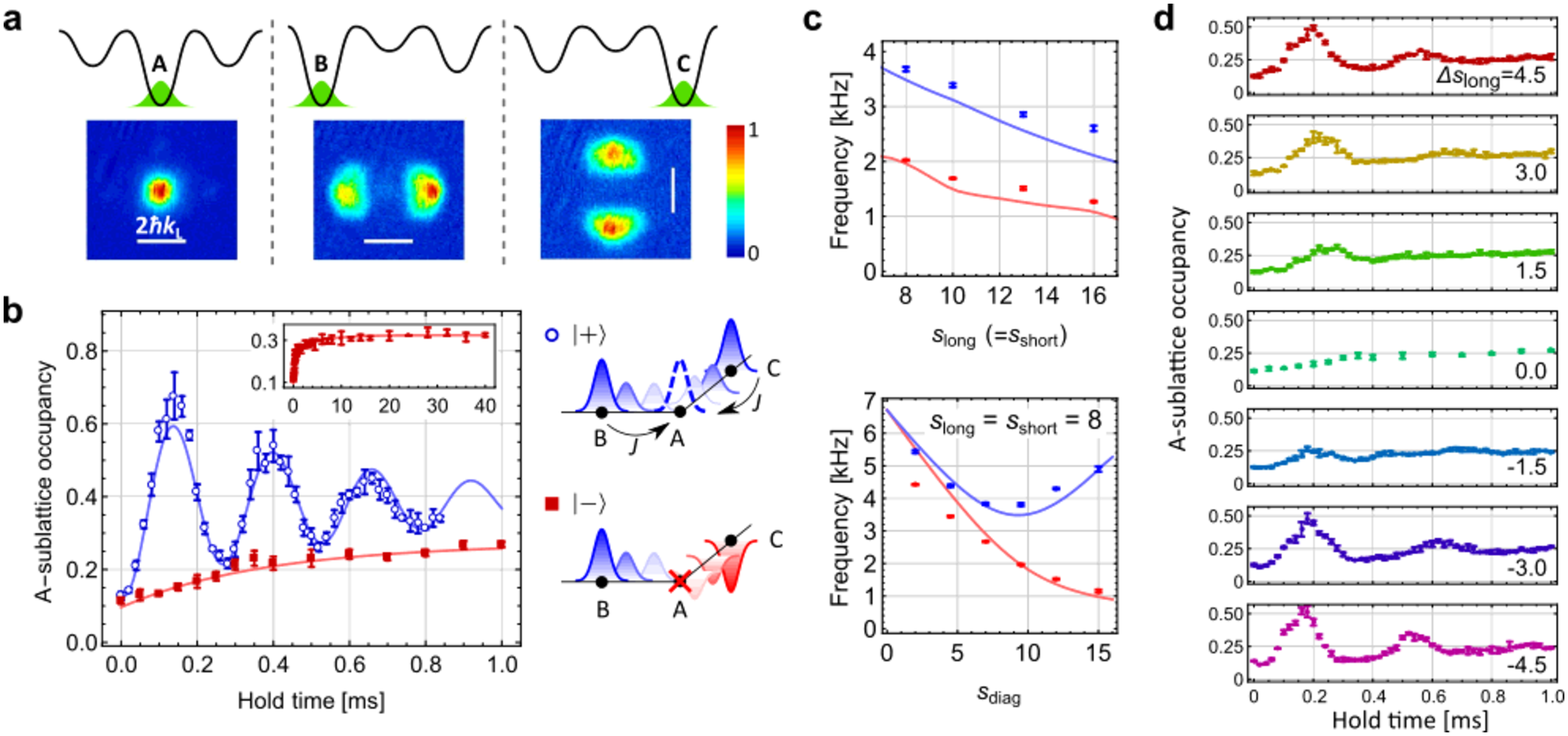}
	\caption{
		\textbf{Tunneling dynamics in the Lieb lattice}.
		\textbf{a}, Demonstrating the measurement of sublattice occupancy. Here, sublattice mapping technique is
		applied to atoms loaded into (left) $((\slong^{(x)},\slong^{(z)}), \sshort, \sdiag) = ((8,8), 8, 0)$,
		(center) $((2,8), 8, 19)$, and (right) $((8,2), 8, 19)$, corresponding to atoms in $A$-, $B$-, and $C$-sites, respectively.
		\textbf{b}, Measured tunneling dynamics of $\ket{+}$ and $\ket{-}$ initial states
		in the Lieb lattice with $(\slong, \sshort, \sdiag) = (8, 8, 9.5)$. Solid lines are the fits to the experimental data
		with damped sinusoidal oscillation (for $\ket{+}$) and double exponentials (for $\ket{-}$).
		Inset shows dynamics of the $\ket{-}$ state for longer hold times. Error bars denote standard deviation.
		Illustration of tunneling process for each initial state is also shown on the right-hand side.
		\textbf{c}, Frequencies of coherent inter-site oscillations. Solid lines are the calculated band gap between
		the 1st and 2nd (red) and the 1st and 3rd bands (blue). Error bars denote fitting error.
		\textbf{d}, Bending flat band. Dynamics of the $\ket{-}$ state in the presence of imbalance
		$\Delta \slong = \slong^{(x)} - \slong^{(z)}$ shows restoration of coherent dynamics.
		Error bars denote standard deviation.
	}
	\label{fig_tunneling}
\end{figure*}

As described above, the most intriguing property of a flat band is the localization of the wave function
at certain sublattice sites.
In the case of the Lieb lattice, the wave function of the flat band vanishes on the $A$-sublattice.
Here we reveal this property by observing the tunneling dynamics of a Bose gas
initially condensed at the $\ket{-}=\ket{B}-\ket{C}$ state, and compare it to the dynamics of the state with
opposite relative phase, $\ket{+}=\ket{B}+\ket{C}$.
In order to observe real-space dynamics of the system, we perform projection measurement of the occupation
number in each sublattice, which we call sublattice mapping. In this method, first we quickly change the lattice potential to
$((\slong^{(x)},\slong^{(z)}), \sshort, \sdiag) = ((8,14), 20, 0)$.
In this configuration, all three sublattices are energetically well separated from one another and the lowest three bands
consist of the $A$-,$B$- and $C$-sublattice, respectively. This maps sublattice occupations to band occupations,
which then can be measured by band mapping technique. Figure \ref{fig_tunneling}a shows the demonstration of
this method, in the cases where atoms occupy only one of the sublattices.
Note that the populations in the $B$- and $C$-sublattices are mapped to the 2nd Brillouin zones for the one-dimensional
lattice along the $x$ and $z$-axis, respectively. This is because the turning off of the diagonal lattice decouples
these two directions and the fundamental bands are labeled by the combination of band indices of 1D lattices.

We prepare the initial state $\ket{+}$ by simply loading a BEC into the Lieb lattice with deep $\vdiag$.
On the other hand, the $\ket{-}$ state is obtained by applying band transfer method to the $\ket{+}$ state.
Dynamics of these initial states after the lattice depths are changed to satisfy the equal-offset condition $E_A = E_B = E_C$
is measured by the sublattice mapping.
As shown in Fig. \ref{fig_tunneling}b, we reveal qualitatively different behaviors of these two states: the $\ket{-}$ state
shows a significant suppression of the $A$-sublattice occupancy, indicating the freezing of the tunneling dynamics to the
$A$-sublattice from the $\ket{-}$ state
with only a slow decay to the $A$-sublattice, whereas the $\ket{+}$ state exhibits coherent oscillations
between the $A$- and ($BC$)-sublattices. This clearly features the geometric structure of the Lieb lattice
mentioned above. Double exponential fit to the data for the $\ket{-}$ initial state yields
$\tau_1=0.36(4)$~ms and $\tau_2=5.5(9)$~ms, indicating that the leakage to the $A$-sublattice is caused by
the decay to the lowest band.

In the Bloch basis, the state $\ket{+}$ is expressed as $\ket{\text{1st}}-\ket{\text{3rd}}$ and its time evolution
is driven by the band gap $\Delta E_{3,1}$ which equals $4\sqrt{2}J$ in the tight binding limit. After a half period
$\pi \hbar/\Delta E_{3,1}$ the state evolves to $\ket{A}=\ket{\text{1st}}+\ket{\text{3rd}}$, leading to
coherent tunneling to the $A$-sublattice.
Similarly, it is possible to arrange the initial lattice depths so that the lowest Bloch state has the maximum overlap with
a certain superposition of  $\ket{\text{1st}}$ and $\ket{\text{2nd}}$. Sudden potential change to the Lieb lattice
drives oscillation between the $B$- and $C$-sublattices, whose frequency gives the band gap $\Delta E_{2,1}$.
We fit these data with a damped sinusoidal oscillation and compare the extracted frequency with the result of
single particle band calculations (see Fig. \ref{fig_tunneling}c). 
Qualitative behavior is well reproduced, while quantitative discrepancies are found.
This is caused by interactions, as we present a systematic study of the density dependence
of the oscillation frequency in Section IV of Supplementary Information.

We further investigate the tunneling dynamics of the $\ket{-}$ initial state by adding the perturbations
which destroy the flatness of the second band. The flatness is robust against the independent change
of nearest-neighbor tunneling amplitudes $J_x$, $J_z$ along the $x$- and $z$-directions, and energy offset $E_{A}$,
just as a dark state in a $\Lambda$-type three level system persists regardless of laser intensities and
detuning from the excited state.
However, if the energy difference between the $B$- and $C$-sublattices is introduced -- the two-photon Raman
off-resonant case --, the flat band is destroyed.
Note that the finite $E_B-E_C$ induces population in the $A$-sublattice even at $k=0$. On the other hand,
the direct diagonal tunneling between the $B$- and $C$-sublattices, which is another flat-band-breaking term
existing in our system, keeps a dark state at $\vect{k}=0$ provided $J_x = J_z$.
We create the energy difference by introducing the imbalance of $\Delta \slong = \slong^{(x)} - \slong^{(z)}$.
Figure \ref{fig_tunneling}d shows the time dependence of the $A$-sublattice population for the $\ket{-}$
initial state. It can be clearly seen that the coherent tunneling dynamics starts to grow as the lattice parameters
deviate from the flat-band condition $\Delta \slong = 0$.

\section{Methods}
	\subsection{Preparation of ${}^{174}$Yb BEC}
		After collecting about $10^7$ atoms with a magneto-optical trap with the intercombination transition,
		the atoms are transferred to a crossed optical trap. Then we perform an evaporative cooling,
		resulting in an almost pure BEC with about $10^5$ atoms with no discernable thermal component.
		
		All of the optical lattice experiments presented in this paper are subject to additional weak confinement
		due to a crossed optical dipole trap operating at $532$~nm.
		Gaussian shape of laser beams for the trap and lattices impose a harmonic confinement on atoms,
		whose frequencies are $(\omega_{x'}, \omega_{y'}, \omega_z)/2\pi = (147, 37, 105)$~Hz at the lattice depths
		of $(\slong, \sshort, \sdiag) = (8, 8, 9.5)$. Here, the $x'$- and $y'$-axes are tilted from the lattice axes ($x$ and $y$)
		by $45^\circ$ in the same plane.

	\subsection{Construction of optical Lieb lattice}
		The relative phases between the long and short lattice ($\phi_x$, $\phi_z$)
		can be adjusted by changing the frequency difference between these lattice beams \cite{Foelling2007}.
		The proper frequencies that realize the Lieb lattice ($\phi_x=\phi_z=0$) are determined by analyzing
		the momentum distribution of a $^{174}$Yb BEC released from the lattice,
		as in the case of the parameter $\psi$ of the diagonal lattice (Supplementary Information).
		The relative phase between the long and short lattices at the position of atoms depends on the optical
		path lengths from common retro-reflection mirrors, and in general two phases $\phi_x$ and $\phi_z$ are not equal.
		We shift the frequency of long lattice laser by an acousto-optic modulator (AOM) inserted in the path for the $z$-axis
		in order to simultaneously realize $\phi_x=\phi_z=0$.
		Optimal frequency difference is sensitive to the alignment of the lattice beams and day-by-day calibration of
		the phases is needed. Typical drift of the required RF frequency for the compensation AOM is within $5$~MHz.

		To stabilize the phase $\psi$, we construct a Michelson interferometer along the optical path of the diagonal lattice
		with frequency stabilized $507$~nm laser. The interferometer has two piezo electric transducer (PZT)-mountded
		mirrors one of which is shared with the lattice laser beam for phase stabilization, and another one for shifting the phase
		over the range $10 \pi$ with stabilization kept active. The short-term stability of $\psi$ is estimated
		to be $\pm 0.007 \pi$.
		The last few optics in front of the chamber are outside of the active stabilization,
		which causes slow drift of $\psi$ due to the change of environment such as temperature.
		The typical phase drift is $0.05 \pi$ per hour, and all measurements of sequential data set are finished
		within $20$ minutes from the last phase calibration.

		At the proper phase parameters $\phi_x=\phi_z=0$ and $\psi=\pi/2$,  the potential depth at the center of
		each site becomes equal when $\vlong=\vshort=\vdiag$. In this condition, however, the energy offset $E_A$
		becomes lower than $E_B$ and $E_C$ because of the difference in the zero-point energies.
		We search optimal $\vdiag$ by single-particle band calculation (see also Supplementary Information for the derivation
		of Hubbard parameters).

	\subsection{Band occupation measurement and sublattice mapping}
		To measure the quasimomentum distribution of atoms, we turn off all the lattice potentials with a exponential form
		\begin{eqnarray}
			V(t) = \left\{
			\begin{array}{cc}
				V(0) \exp (-4t/\tau)	&	0<t<\tau\\
				0	&	t \geq \tau
			\end{array}
			\right.
		\end{eqnarray}
		with a time constant $\tau=0.6$~ms.
		The dipole trap is kept constant during band mapping for preventing movement of the trap center due to gravity
		and suddenly turned off at $t=\tau$. Because of the relatively heavy mass of Yb, the existence of harmonic confinement
		imposes severe restriction on the choice of mapping time $\tau$. We find $\tau > 1$~ms causes considerable deformation
		of the distribution whereas $\tau > 1.5$~ms is desirable to suppress interband transition.
		Due to this non-adiabaticity, up to $20$\% of atoms occupying a certain Brillouin zone are detected in
		its neighboring zones, depending on the shape of the observed quasimomentum distribution.

\section{Acknowledgements}
	We thank K. Noda, K. Inaba, M. Yamashita, I. Danshita, S, Tsuchiya, C. Sato, S. Capponi, Z. Wei and Q. Zhou
	for valuable discussions.
	This work was supported by the Grant-in-Aid for Scientific Research of JSPS (No. 25220711) and
	the Impulsing Paradigm Change through Disruptive Technologies (ImPACT) program.

\clearpage
\setcounter{figure}{0}
\begin{center}
	{\large \textbf{Supplementary Information}}
\end{center}
\renewcommand{\figurename}{FIG. S}
\newcommand{\figureref}[1]{S\ref{#1}}
\section{Calibration of the relative phase}
\begin{figure*}[bt]
	\includegraphics[width=170mm]{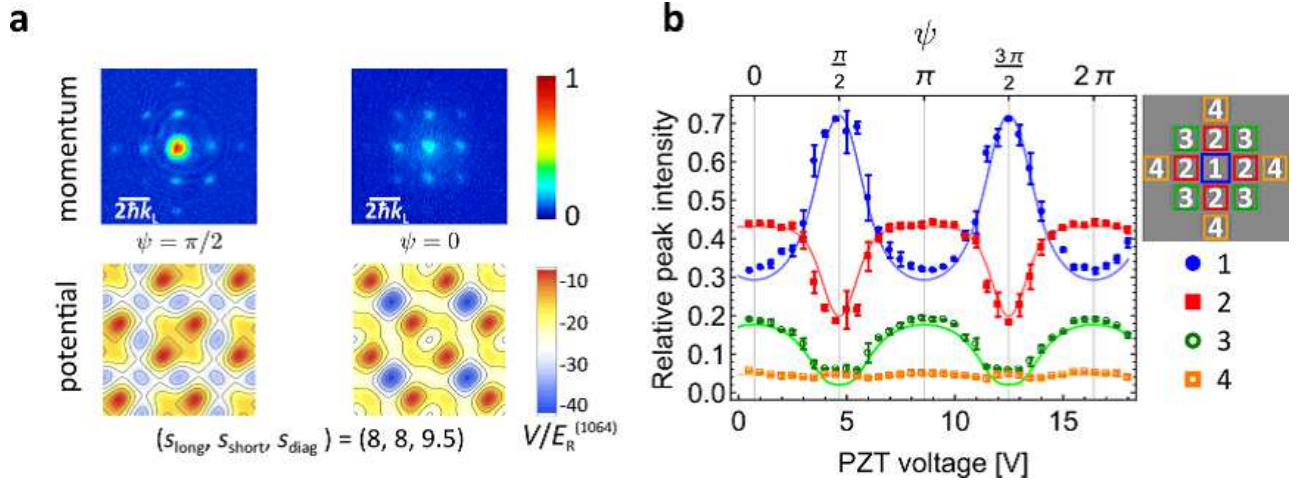}
	\caption{
		\textbf{Phase dependence of a time-of-flight signal.}
		\textbf{a}, Absorption images of a $^{174}$Yb BEC released from the optical lattice
		with $\psi=\pi/2$ (left) and $\psi=0$ (right), taken after $14$~ms of ballistic expansion.
		Corresponding lattice potentials are also shown.
		\textbf{b}, Intensities of the coherent momentum peaks of a BEC as a function of the phase $\psi$,
		measured by the time-of-flight experiment. Each peak is categorized into four, as shown in the right
		of the data plot. Solid lines show the result of single particle band calculation,
		for which horizontal offset and scale are adjusted to match the experimental data.
		Error bars denote standard deviation of three independent measurements.
	}
	\label{fig_phase}
\end{figure*}

Upper images in Fig.~\figureref{fig_phase}a shows time-of-flight absorption images of a $^{174}$Yb BEC loaded into
an optical lattice at the phase $\psi = \pi/2$ and $\psi =0$, leading to lattice potential landscapes
shown in the lower images.
In the case of $\psi=\pi/2$, the ground state wave function distributes over all three sites in a unit cell.
On the other hand, for $\psi=0$ atoms are strongly localized on a square lattice composed of $A$-sites.
This feature is reflected in a momentum distribution of a BEC, where the fraction of zero momentum coherent peak
becomes much larger for $\psi=\pi/2$ due to the delocalization.

The phase dependence of the each momentum peak of a BEC is shown in Fig.~\figureref{fig_phase}b.
The experimental data are well reproduced by the non-interacting band calculation,
which is utilized for the calibration of $\psi$.

\section{Tight-binding model for the optical Lieb lattice}
\begin{figure}[bt]
	\includegraphics[width=85mm]{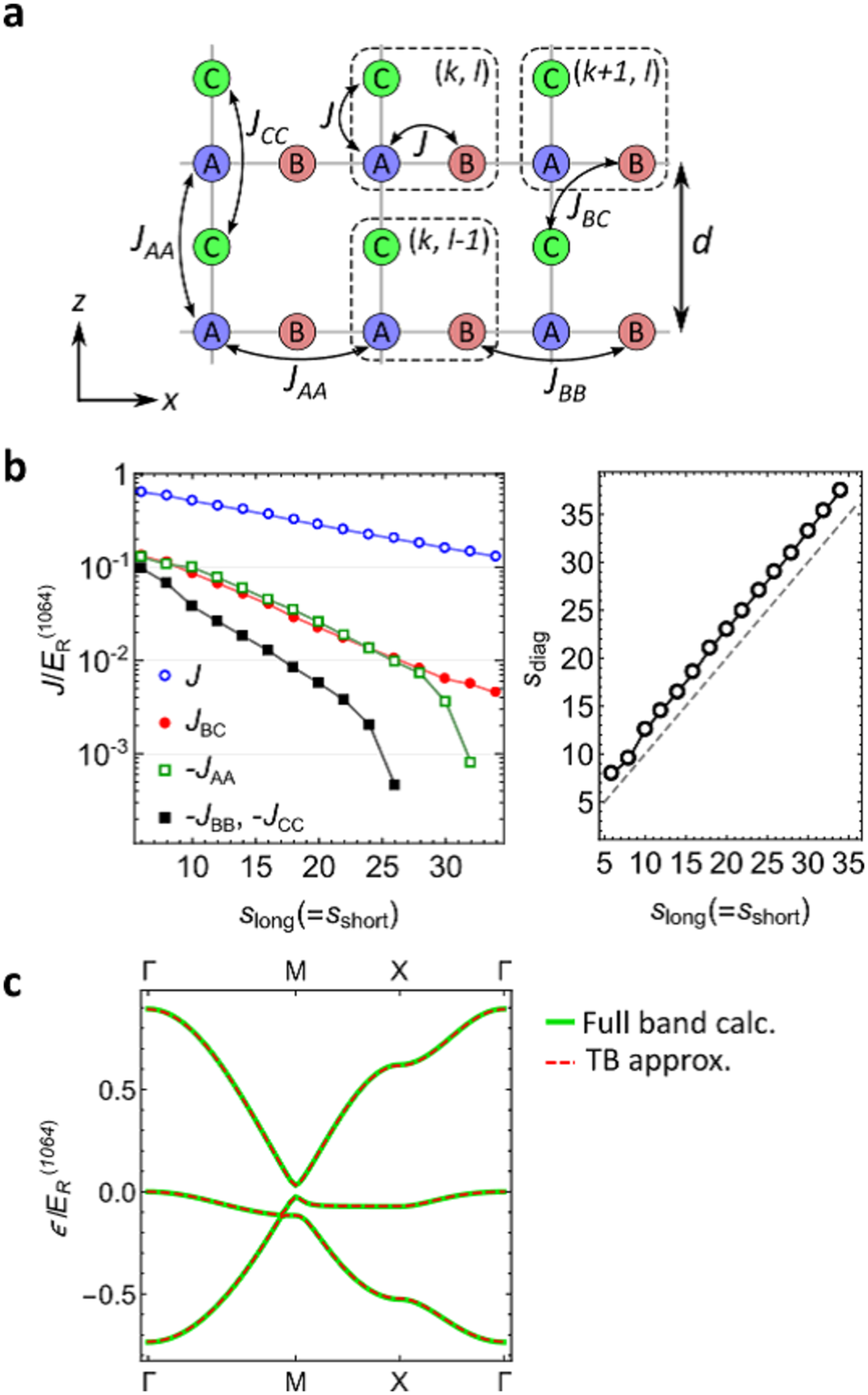}
	\caption{
		\textbf{Tunneling parameters in the optical Lieb lattice.}
		\textbf{a}, Tight binding model for the Lieb lattice with beyond-nearest-neighbor hopping.
		\textbf{b}, (left)
		Tunneling amplitudes obtained from the tight-binding approximation are shown as a function of lattice depth.
		The solid lines are linear interpolations for the successive data points.
		Signs of $J_{AA}$, $J_{BB}$ ($J_{CC}$) are inverted to display in a log scale.
		(right) Selected diagonal lattice depth $\sdiag$ to satisfy $E_A=E_B=E_C$.
		The dashed line represents $\sdiag=\slong (=\sshort)$ for reference.
		\textbf{c}, Band structure of the Lieb lattice with $(\slong, \sshort,\sdiag) = (20, 20, 23)$.
		Dispersion relations of the lowest three bands of the first-principle band calculation (green solid)
		are well reproduced by the tight binding approximation presented here. The best fit hopping parameters
		are $J=0.282$, $J_{BC}=0.022$, $J_{AA}=-0.026$, $J_{BB}=J_{CC}=0.0057$, $E_A=0.140$ and
		$E_B = E_C=-0.058$, in unit of $\er$. The origin of the energy is set to the 2nd band at the $\Gamma$ point.
	}
	\label{fig_tunneling2}
\end{figure}

Here we present the derivation of the tight-binding model describing our optical Lieb lattice.
The kinetic energy part of the tight-binding Hamiltonian considered here is written in the form
\begin{equation}
\hat{H}_{\rm TB} = \hat{H}_{\rm Lieb} + \hat{H}_{\rm BC} + \hat{H}_{\rm S}.
\end{equation}
Here $\hat{H}_{\rm Lieb}$ represents the part involving nearest-neighbor hopping,
$\hat{H}_{\rm BC}$ next-nearest-neighbor hopping between the $B$ and $C$ sites,
and $\hat{H}_{\rm S}$ the Hamiltonian within each sublattice.
These terms are explicitly expressed in the second quantized form as
\begin{widetext}
	\begin{eqnarray}
	\hat{H}_{\rm Lieb} &=& -J\sum_{k,l}
	\left[
	\left( \hat{a}^{\dagger}_{(k,l),A} \hat{a}_{(k,l),B} + \hat{a}^{\dagger}_{(k,l),A} \hat{a}_{(k-1,l),B} \right) 
	+\left( \hat{a}^{\dagger}_{(k,l),A} \hat{a}_{(k,l),C} + \hat{a}^{\dagger}_{(k,l),A} \hat{a}_{(k,l-1),C} \right)
	\right]  + \text{H.c.},\\
	\hat{H}_{\rm BC} &=& -J_{BC} \sum_{k,l}
	\left[
	\left( \hat{a}^{\dagger}_{(k,l),B} \hat{a}_{(k+1,l),C} + \hat{a}^{\dagger}_{(k,l),B} \hat{a}_{(k,l-1),C} \right)
	\right]  + \text{H.c.}, \\
	\hat{H}_{\rm S} &=& \sum_{k,l}
	\left[
	-\sum_{S=A,B} J_{SS} \left( 
	\hat{a}^{\dagger}_{(k,l),S} \hat{a}_{(k+1,l),S} + \text{H.c.} \right)
	-\sum_{S=A,C}  J_{SS} \left(\hat{a}^{\dagger}_{(k,l),S} \hat{a}_{(k,l+1),S} + \text{H.c.} \right) \right.
	\nonumber \\
	&+& \left. \sum_{S=A,B,C} E_{S} \hat{a}^{\dagger}_{(k,l),S} \hat{a}_{(k,l),S} \right]
	\end{eqnarray}
\end{widetext}
where $\hat{a}_{(k,l),S}$ is the annihilation operator on a site $S$ ($=A, B, C$) in a unit cell labeled by its coordinates
$(x,z) = (kd, ld)$.
Inclusion of beyond-nearest-neighbor hopping $J_{BC}$ and $J_{SS}$ is necessary to
reproduce the band dispersions obtained by the first principle band calculations, especially for shallow lattices.
Figure~\figureref{fig_tunneling2}a. shows a sketch of each hopping term.
Regarding $J_{BC}$, we consider terms along the direction $\hat{\vect{x}}+\hat{\vect{z}}$ only,
because our diagonal lattice suppresses the hopping along the $\hat{\vect{x}}-\hat{\vect{z}}$ direction
(see Fig.~\figureref{fig_phase}a). Similarly, hopping $J_{BB}$ ($J_{CC}$) is restricted to the $x$- ($z$-) direction,
respectively.
In the momentum space representation
$
\hat{a}_{i,S} = \frac{1}{\sqrt{N}} \sum_{\svect{k}}
\ex^{\im \svect{k} \cdot \svect{x}_{i, S}} \hat{a}_{\svect{k},S},
$
the Hamiltonian is diagonalized with respect to the momentum indices as
\begin{eqnarray}
&\hat{H}_{\rm TB} = \sum_{\svect{k}}
\left(
\begin{array}{ccc}
\hat{a}^{\dagger}_{\svect{k},A}	&	\hat{a}^{\dagger}_{\svect{k},B}	&	\hat{a}^{\dagger}_{\svect{k},C}\\
\end{array}
\right) 
\cal{T}
\left(
\begin{array}{c}
\hat{a}_{\svect{k},A}	\\
\hat{a}_{\svect{k},B}	\\
\hat{a}_{\svect{k},C}	\\
\end{array}
\right), \\
&\cal{T} = \cal{T}_{\rm Lieb} + \cal{T}_{\rm BC} + \cal{T}_{\rm S}.
\end{eqnarray}
Here $\cal{T}$ is the $3\times 3$ matrix which couples each sublattice, given by
\begin{widetext}
	\begin{eqnarray}
	&\cal{T}_{\rm Lieb} = 
	\left(
	\begin{array}{ccc} 
	0	&	-2J \cos (k_x d/2)	&	-2J \cos (k_z d/2)	\\
	-2J \cos (k_x d/2)	&	0	&	0	\\
	-2J \cos (k_z d/2) 	&	0	&	0	\\	
	\end{array}
	\right), \nonumber \\
	&\cal{T}_{\rm BC} = 
	\left(
	\begin{array}{ccc} 
	0	&	0 	&	0	\\
	0	&	0 	&	-2J_{BC} \cos (k_x d/2 + k_z d/2) \\
	0 	&	-2J_{BC} \cos (k_x d/2 + k_z d/2) &	0\\	
	\end{array}
	\right), \nonumber \\
	&\cal{T}_{\rm S} = 
	\left(
	\begin{array}{ccc}
	E_{A} -2J_{AA} \left[ \cos (k_x d) + \cos (k_z d) \right]	&	0	&	0	\\
	0	&	E_{B} -2J_{BB} \cos (k_x d)	&	0	\\
	0	&	0	&	E_{C} -2J_{CC}  \cos (k_z d)	\\
	\end{array}
	\right) .	\label{Tmatrix}
	\end{eqnarray}
\end{widetext}
Leaving only $\cal{T}_{\rm Lieb}$, we have analytic expressions for the eigenvalues of the Lieb lattce
with hopping to the nearest neighbors only, as
\begin{eqnarray}
E_{\pm} &=& \pm 2J \sqrt{\cos^2 (k_xd/2) + \cos^2 (k_zd/2)}, \\
E_0 &=& 0,
\end{eqnarray}
and the corresponding eigenfunctions
\begin{eqnarray}
\ket{\vect{k},\text{1st}} &=& \frac{1}{\sqrt{2}} \left( \ket{\vect{k},A} + \sin \theta_{\svect{k}} \ket{\vect{k},B}
+ \cos \theta_{\svect{k}} \ket{\vect{k},C} \right), \qquad\\
\ket{\vect{k},\text{2nd}} &=&
\cos \theta_{\svect{k}}\ket{\vect{k},B}- \sin \theta_{\svect{k}} \ket{\vect{k},C}, \qquad\\
\ket{\vect{k},\text{3rd}} &=& \frac{1}{\sqrt{2}} \left( \ket{\vect{k},A} - \sin \theta_{\svect{k}} \ket{\vect{k},B}
- \cos \theta_{\svect{k}} \ket{\vect{k},C} \right), \qquad
\end{eqnarray}
with $\tan \theta_{\svect{k}} = \cos(k_xd/2)/\cos(k_zd/2)$.
To obtain experimentally relevant tight-binding parameters, we perform least-square fitting of the band dispersion
obtained by diagonalizing $\cal{T}$ to the lowest three bands of the first principle calculations. 
This procedure gives the optimal hopping amplitudes which reproduce the actual band structures,
as shown in Figures~\figureref{fig_tunneling2}b and c. Dominant contribution other than the nearest neighbor
hopping $J$ comes from $J_{BC}$ and $J_{AA}$. While the latter does not affect the flat band,
$J_{BC}$ eliminates the dark states and causes finite dispersion of the flat band. The ratio $J_{BC}/J$ can be
made smaller by increasing lattice depth, or adding another diagonal lattice along the
$\hat{\vect{x}}+\hat{\vect{z}}$ direction.

\begin{figure}[bt]
	\includegraphics[width=85mm]{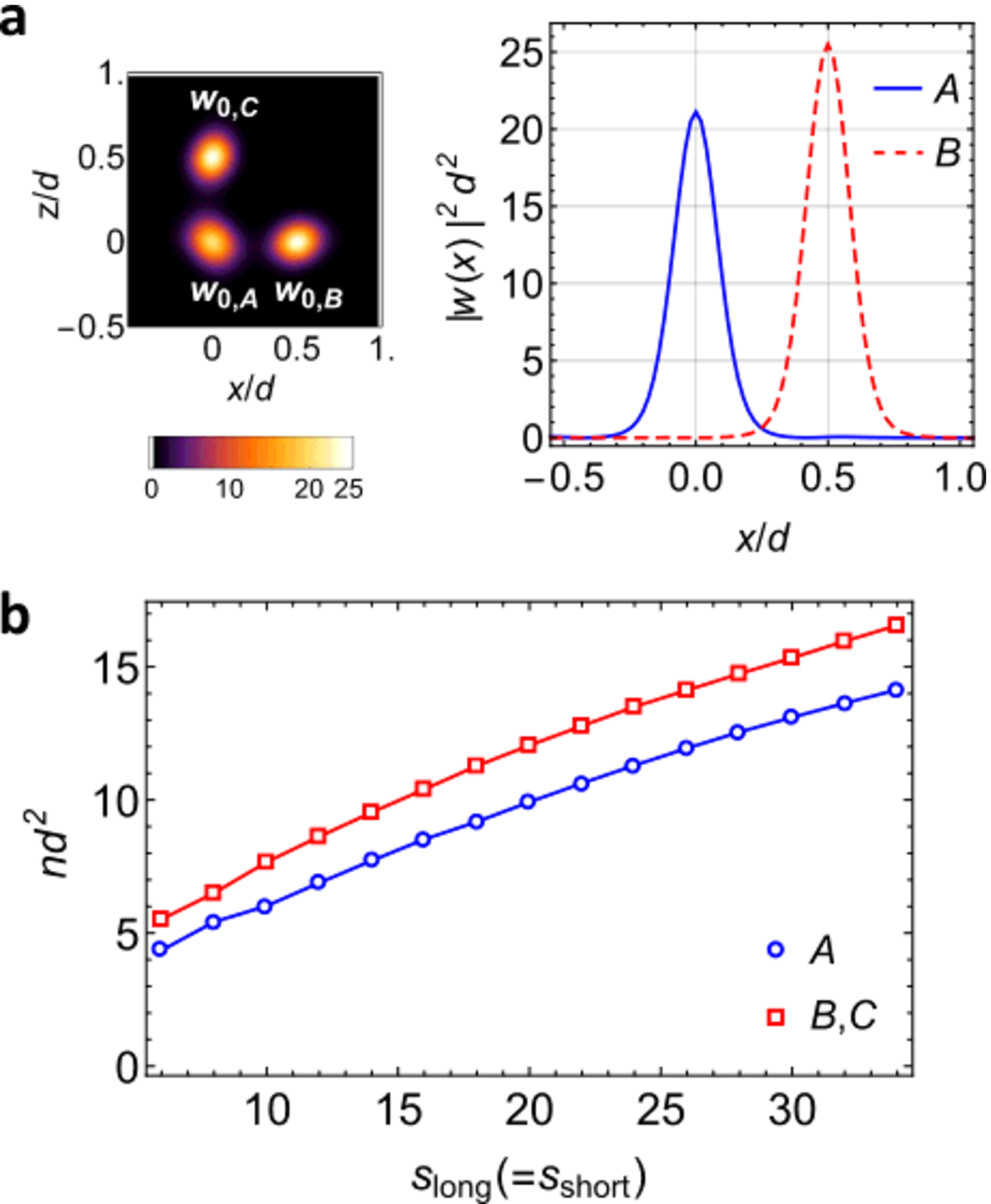}
	\caption{
		\textbf{Wannier functions of the optical Lieb lattice.}
		\textbf{a}, Wannier functions of the Lieb lattice composed of the lowest three bands, at the lattice depths
		of $(\slong, \sshort,\sdiag) = (20, 20, 23)$. Scaled densities $|w_{0,S}(x,z)|^2 \times d^2$ are shown.
		Density profiles of the Wannier functions on $A$ and $B$ sites along the $z=0$ line are also shown
		in the right hand side.
		\textbf{b}, 
		Average densities of the 2D Wannier functions of the Lieb lattice are plotted as a function of lattice depth.
		The solid lines are linear interpolations for the successive data points.
	}
	\label{fig_wannier}
\end{figure}

To derive parameters such as on-site interactions, the above procedure is not sufficient and we need to
construct Wannier functions. This involves ambiguity in the definition of the phase of each Bloch state,
and we should choose the phase such that resulting Wannier functions are well localized and
minimize non-Hubbard type interactions.
Let $(u_{\svect{k},A}^{(n)}, u_{\svect{k},B}^{(n)}, u_{\svect{k},C}^{(n)})$ ($n=1,2,3$) to be
the eigenvectors of $\cal{T}$ with the $n$-th energy band.
The Bloch states obtained from the band calculations $\ket{\vect{k},n}$ will be written as superposition of
the sublattice momentum eigenstates $\ket{\vect{k},S}= \hat{a}^{\dagger}_{\svect{k},S} \ket{0}$, in the form
\begin{eqnarray}
\ex^{\im \theta (\svect{k},n)} \ket{\vect{k},n} = \sum_{S=A,B,C} u_{\svect{k},S}^{(n)} \ket{\vect{k},S}.
\end{eqnarray}
Here $\theta (\vect{k},n)$ is unknown phase mentioned above.
We follow the procedure similar to that described in \cite{Luhmann2014} and choose the phase $\theta$ so that
all Bloch states constructively interfere at the specific lattice site to give the localized Wannier function on that site:
\begin{eqnarray}
\theta(\vect{k},n) = \arg \left[ \ex^{-\im \svect{k}\cdot \svect{x}_{i,S_{\rm ref}}} u_{\svect{k},S_{\rm ref}}^{(n)}
\psi_{\svect{k}}^{(n)} (\vect{x}_{i, S_{\rm ref}}) \right]
\end{eqnarray}
where $\psi_{\svect{k}}^{(n)} (\vect{x})$ is the Bloch wave function $\langle \vect{x}|\vect{k},n \rangle$.
According to Bloch's theorem, the above expression actually does not depend on the choice of unit cell $i=(k,l)$.
For each $\vect{k}$ and $n$, we choose appropriate reference sublattice $S_{\rm ref}$ where the Bloch state
has the largest amplitude.
After all, the Wanner functions $w_{i,S}(\vect{x})$ can be obtained from the Bloch states as
\begin{eqnarray}
w_{i,S}(\vect{x}) = \sum_{\svect{k}} \ex^{-\im \svect{k}\cdot \svect{x}_{i,S}} 
\sum_{n=1}^{3} u_{\svect{k},S}^{(n)} \ex^{-\im \theta(\svect{k},n)} \psi_{\svect{k}}^{(n)} (\vect{x}).
\end{eqnarray}
Figure~\figureref{fig_wannier}a shows the calculated Wannier functions of the optical Lieb lattice derived by
the above procedure. It can be seen that the three Wannier functions $w_{0,A}$, $w_{0,B}$ and $w_{0,C}$
are well localized at the site $A$, $B$ and $C$ of the unit cell $i=(0,0)$, respectively. It should be noted that,
due to the difference in the confinement of potential wells, density at the $(BC)$-sublattice becomes
higher than that of the $A$-sublattice.
On-site interactions in the Hubbard model is proportional to the two-dimensional average density
$n=\int {\rm d}x{\rm dz} |w_{i,{\rm S}}(\vect{x})|^4$ of the Wannier functions.
As shown in Fig.~\figureref{fig_wannier}b, the sublattice dependence of $n$ amounts to $\sim 20\%$ for our lattice.
\section{Momentum distributions in coherent band transfer}
\begin{figure}[tb]
	\includegraphics[width=85mm]{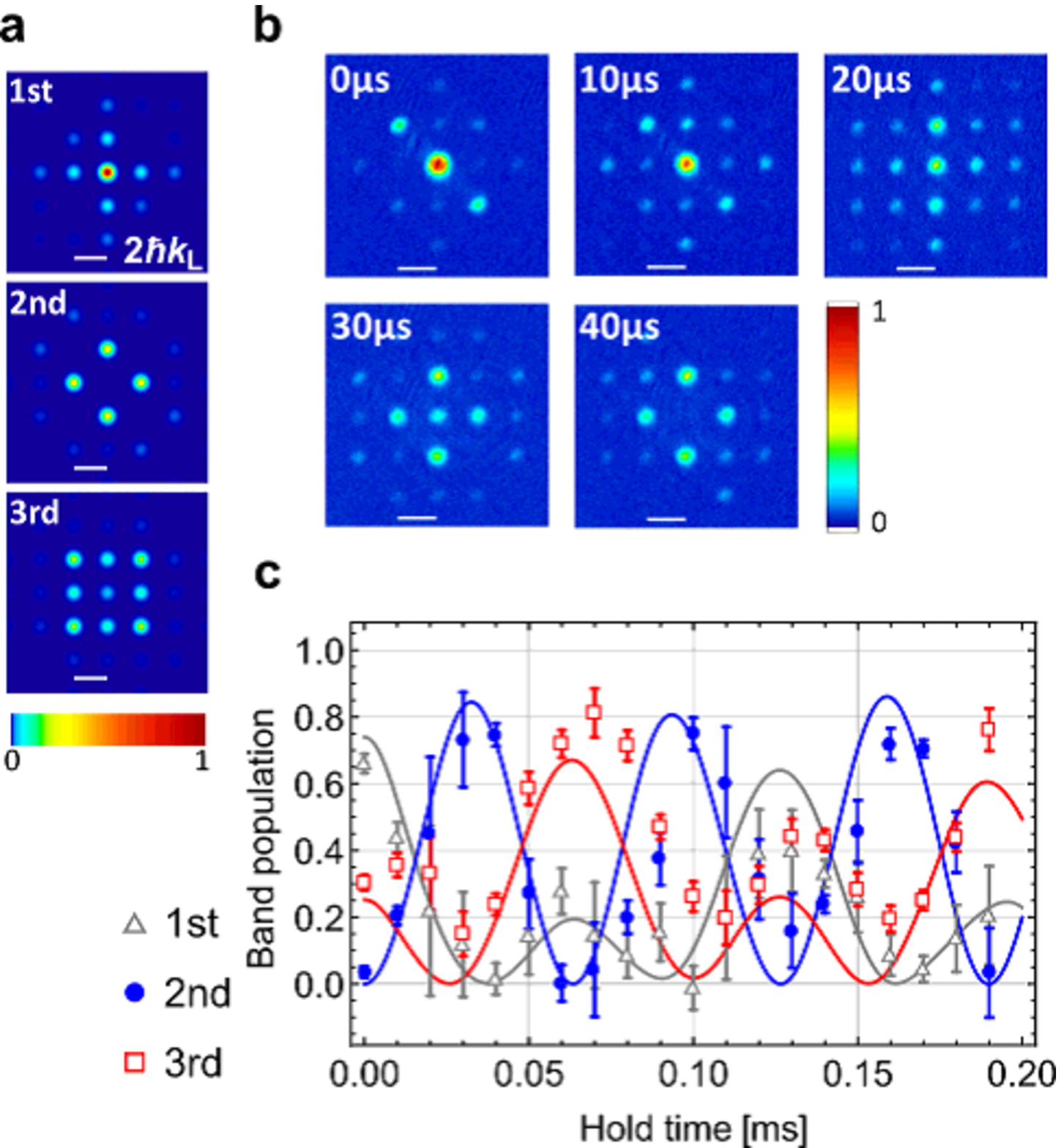}
	\caption{
		\textbf{Momentum space observation of coherent band transfer.}
		\textbf{a}, Numerically calculated momentum distributions of the zero-momentum eigenstates
		of the lowest three bands. Potential depths are set to $(\slong, \sshort, \sdiag) = (8, 8, 9.5)$.
		\textbf{b}, Evolution of the momentum distribution of a BEC during coherent band transfer.
		Each absorption image is taken after $14$~ms of ballistic expansion.
		\textbf{c}, Oscillations of the band populations during coherent band transfer.
		Each population is obtained by fitting time-of-flight images with the momentum distribution
		of the superposition of the lowest three bands.
		Solid lines are theoretical expectation, using the parameters obtained by fitting to the band
		mapping measurement in Fig.~2c of the main paper.
		Error bars denote standard errors in fitting procedure.
	}
	\label{fig_momentum}
\end{figure}

In the main paper, coherent transfer of a ground state BEC into the 2nd band of the Lieb lattice
is analyzed with the quasimomentum distributions. Here, we show a bare momentum distribution in
the transfer process, measured by direct time-of-flight experiment.
As mentioned in the main paper, quasimomentum analysis cannot distinguish a condensate in the center of the 2nd
band from that in the 3rd band. In contrast, these two states have quite different distributions in the momentum space, as shown in Fig.~\figureref{fig_momentum}a. The 2nd band is characterized by the prominent peaks at
$\vect{k} = (\pm 2\kl, 0)$ and $(0, \pm 2\kl)$, and vanishing central peak due to the destructive
interference of the wave function on the $B$- and $C$-sublattices. The 3rd band is distinguished from the other
bands by its large population at $\vect{k} = (\pm 2\kl, \pm 2\kl)$.

Figure~\figureref{fig_momentum}b shows the measured evolution of the momentum distribution of a BEC
during the coherent transfer scheme. The initial state dramatically changes its distributions and
the feature of the 2nd band is well reproduced around  $\sim 40$~$\mu$s. 
At each evolution time, the wave function of a BEC is well described by a certain superposition of
the three eigenstates as
\begin{eqnarray}
\ket{\psi} = c_1 \ket{\text{1st}}
+ c_2 \ex^{\im \alpha} \ket{\text{2nd}} + c_3 \ex^{\im \beta} \ket{\text{3rd}},
\end{eqnarray}
where the parameters $c_1$, $c_2$, $c_3$, $\alpha$ and $\beta$ are chosen to be real and
satisfy $\sum_{i=1}^{3}c_{i}^{2}=1$. Since the basis states $\ket{\text{1st}}$, $\ket{\text{2nd}}$
and $\ket{\text{3rd}}$ are far from the eigenstates during the transfer process, these parameters
show complicated time evolutions. By regarding them as free parameters, we fit the experimental
data with the distribution $\left| \langle \vect{k}|\psi(t) \rangle \right|^2$ to obtain the band populations
$c_{i}^2$ ($i=1,2,3$). In fitting, relative intensities of the coherent peaks in the region
$-4\kl \leq k_x \leq4\kl$ and $-4\kl \leq k_z \leq4\kl$ are used.
Obtained evolutions of the band populations are plotted in Fig.~\figureref{fig_momentum}c, which confirms
the efficient transfer into the 2nd band and show agreement with band mapping measurement in the main paper.
Rather scattered data points might be due to the fitting to small coherent peaks whose
intensities are largely affected by the background noise, or the systematic contribution from the
higher energy bands. 

\section{Effect of interactions on inter-sublattice oscillations of a BEC}
\begin{figure*}[tb]
	\includegraphics[width=170mm]{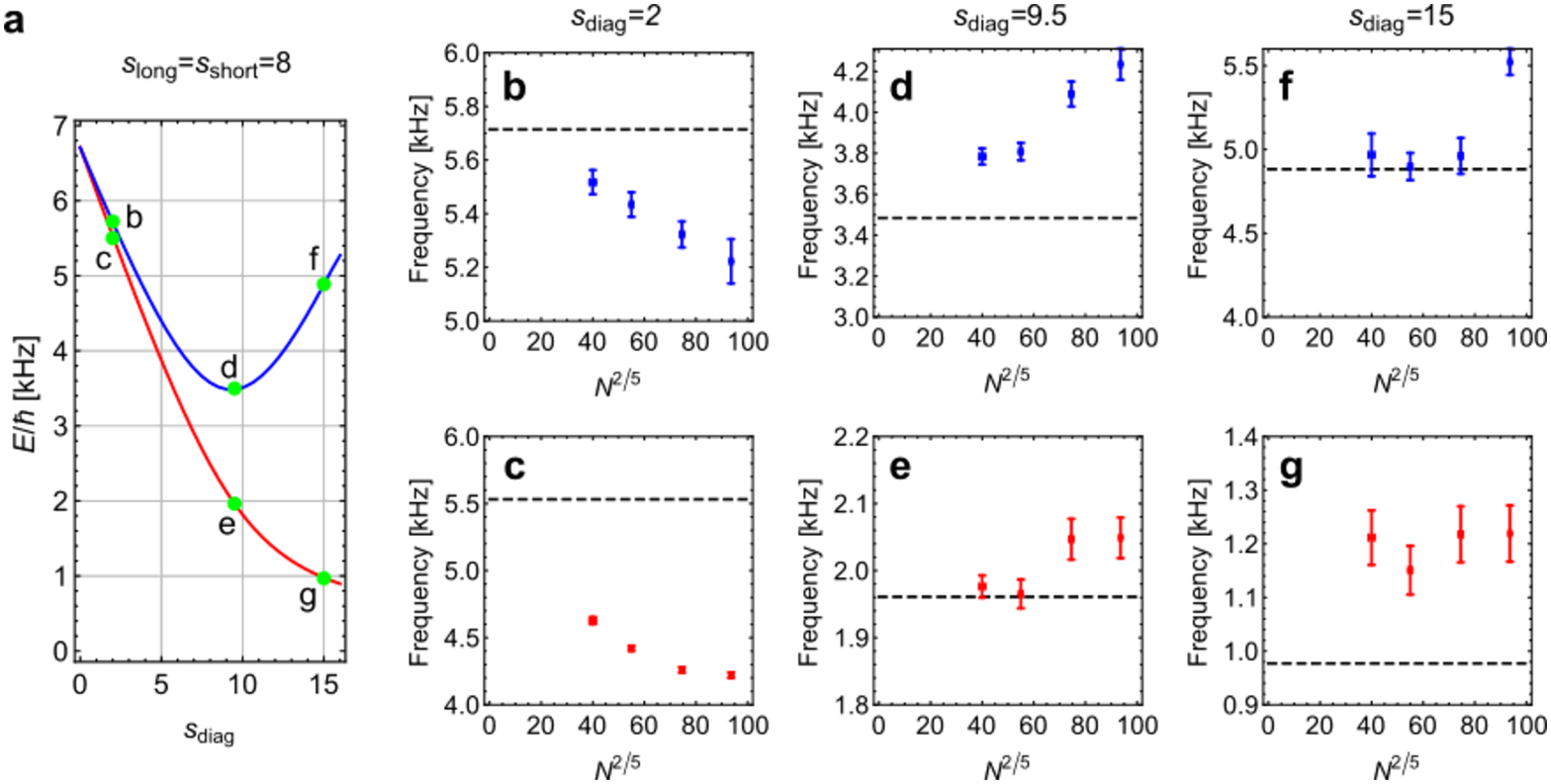}
	\caption{
		\textbf{Density dependence of oscillation frequency.}
		\textbf{a}, Numerically calculated band structures of the optical Lieb lattice, as a function of $\sdiag$.
		Green circles are the points where the measurements in \textbf{b-g} are performed,
		with corresponding alphabet symbols.
		\textbf{b-g}, Density dependence of inter-sublattice oscillation frequency.
		Horizontal dashed lines indicate the band gap from single particle calculations shown in \textbf{a}.
		Error bars denote standard errors in fitting procedure.
	}
	\label{fig_bandgap}
\end{figure*}

In Fig.~4c of the main paper, we compare the measured frequencies of inter-sublattice oscillations
of a BEC with the relevant band gaps obtained by single particle band calculations.
Systematic deviation from the prediction is found especially in the regime of small $\sdiag$.
We experimentally examine the effect of interactions on the oscillation frequencies by changing
total atom number of a BEC, at several diagonal lattice depths shown in Fig.~\figureref{fig_bandgap}a.

Assuming that the system is locally uniform, introduction of a local chemical potential
$\mu(\vect{r}) = \mu-V(\vect{r})$ leads to the expression of the total atom number
$N=\int {\rm d}^3r\, n(\mu(\vect{r}))$, where $V(\vect{r})$ is an external harmonic confinement
and $n(\mu)$ is the density averaged over a unit cell.
For a uniform, weakly interacting BEC, the chemical potential has a linear dependence on
atomic density, leading to $N^{2/5}$ dependence of the central density $n(0)$.
Therefore, we plot the observed oscillation frequency as a function of $N^{2/5}$ in Fig.~\figureref{fig_bandgap}.
For each case, an extrapolation to the zero density seems to reproduce the expected band gap.
Remaining errors of $\sim 200$~Hz can be accounted by uncertainty in the calibration of the lattice depths.
Systematic errors in the oscillation measurements can also come from finite lifetime ($\sim 1$~ms) of
a condensate in the excited bands, which matters especially in the low frequency modes.

Whereas the negative shifts of the oscillation frequencies are observed in small $\sdiag$,
the shifts turn into positive around the Lieb lattice condition $\sdiag=9.5$. This tendency can be
qualitatively explained by considering the density distribution of each band.
For small $\sdiag$, $E_A$ becomes much lower than $E_B$ and $E_C$ and the wave function of
the lowest band concentrates on the $A$-sublattice, whereas the 2nd and 3rd bands have population
in both the $B$- and $C$-sublattices. The existence of a repulsive interaction shifts the energy upwards
by a greater amount for the lowest band, leading to the reduction of the gaps to the higher bands.
On the other hand, at $\sdiag \sim 9.5$ the ground state wave function spreads over all sublattices and
the amount of energy shift will decrease. The oscillations frequencies may depend on the initial preparation,
i.e., the fraction of each eigenstate.
The initial conditions for the measurement in Fig.~\figureref{fig_bandgap} are listed in Table~\ref{tbl_depth}.

\begin{table*}[tb]
	\begin{tabular}{l|l|ccc}
		$s_{\rm meas}$ $[\slong, \sshort, \sdiag]$	&	$s_{\rm init}$ $[((\slong^{(x)},\slong^{(z)}), (\sshort^{(x)},\sshort^{(z)}), \sdiag]$	&	$\left| \langle 1,\text{meas}|\text{init}\rangle \right|^2$	&	$\left| \langle 2,\text{meas}|\text{init}\rangle \right|^2$	&	$\left| \langle 3,\text{meas}|\text{init}\rangle \right|^2$	\\
		1st-2nd		&	&	&	&	\\
		(8, 8, 2)	&	((7.6, 0), (8.7, 29), 6)	&	0.964	&	0.033	&	$<10^{-6}$	\\
		(8, 8, 9.5)	&	((7.5, 0), (5.9, 29), 8.4)	&	0.723	&	0.275	&	$<10^{-6}$	\\
		(8, 8, 15)	&	((7.8, 0.03), (3.81, 29), 10)	&	0.571	&	0.426	&	$<10^{-5}$	\\ \hline
		1st-3rd		&	&	&	&	\\
		(8, 8, 2)	&	((8, 8), (8, 8), 20)	&	0.270	&	0	&	0.686	\\
		(8, 8, 9.5)	&	((8, 8), (8, 8), 20)	&	0.739	&	0	&	0.253	\\
		(8, 8, 15)	&	((8, 8), (8, 8), 20)	&	0.977	&	0	&	0.022	\\
	\end{tabular}
	\caption{
		Initial conditions for the inter-sublattice oscillations. For each lattice depths $s_{\rm meas}$ at which oscillations
		are observed, we prepare appropriate initial state by loading a BEC into the lattice with depths $s_{\rm init}$, and
		drive oscillations by suddenly change the depth to $s_{\rm meas}$. Calculated overlap between the initial wave function
		$\ket{\text{init}}$ and each zero-momentum eigenstate of the measurement stage
		$\ket{n, \text{meas}}$ ($n=1,2,3$) is also shown.
	}
	\label{tbl_depth}
\end{table*}

\end{document}